\documentclass[
aps,
prl,
twocolumn,
superscriptaddress]{revtex4-2}
\usepackage{amssymb}
\usepackage{amsmath}
\usepackage{amsthm}
\usepackage[english]{babel}
\usepackage{bm}
\usepackage{bbold}
\usepackage{hyperref}
\usepackage{graphicx}
\usepackage{subfigure}
\usepackage{xcolor}
\usepackage{enumitem}
\usepackage{blindtext}
\newcommand{\ket}[1]{|#1\rangle}
\newcommand{\bra}[1]{\langle #1|}
\newcommand{\inp}[2]{\langle #1|#2\rangle}

\def\<{\langle}  
\def\>{\rangle}  

\newtheorem{theorem}{Theorem}

\hypersetup{
	colorlinks  = true,
	urlcolor    = black,
	citecolor   = blue,
	linkcolor   = blue
}

\DeclareMathSizes{11}{9}{7}{5}

\usepackage[normalem]{ulem}

\begin{document}
	\title{Minimal trade-off and optimal measurement for multiparameter quantum estimation}
	\author{Lingna Wang}
	\email{lnwang@mae.cuhk.edu.hk}
	\affiliation{Department of Mechanical and Automation Engineering, The Chinese University of Hong Kong, Shatin, Hong Kong SAR, China}
	\affiliation{The Hong Kong Institute of Quantum Information Science and Technology, The Chinese University of Hong Kong, Shatin, Hong Kong SAR, China}
    \affiliation{State Key Laboratory of Quantum Information Technologies and Materials, The Chinese University of Hong Kong, Shatin, Hong Kong SAR, China}
 	\author{Hongzhen Chen}
    \email{hzchen@szu.edu.cn}
    \affiliation{Institute of Quantum Precision Measurement, State Key Laboratory of Radio Frequency Heterogeneous Integration, College of Physics and Optoelectronic Engineering, Shenzhen University, Shenzhen, China}
	\author{Haidong Yuan}
	\email{hdyuan@mae.cuhk.edu.hk}
	\affiliation{Department of Mechanical and Automation Engineering, The Chinese University of Hong Kong, Shatin, Hong Kong SAR, China}
	\affiliation{The Hong Kong Institute of Quantum Information Science and Technology, The Chinese University of Hong Kong, Shatin, Hong Kong SAR, China}
    \affiliation{State Key Laboratory of Quantum Information Technologies and Materials, The Chinese University of Hong Kong, Shatin, Hong Kong SAR, China}
	
	\date{\today}
	
	\begin{abstract}
A fundamental challenge in multiparameter quantum estimation arises from the incompatibility of optimal measurements for different parameters, leading to intricate precision trade-offs that obscure the understanding of ultimate quantum limits. Here, we present an approach that precisely quantifies these trade-offs for an arbitrary number of parameters encoded in pure quantum states. Our approach not only derives tight analytical bounds for the trade-offs induced by measurement incompatibility but also provides a systematic methodology to design optimal measurement strategies that saturate these limits. To demonstrate the practical significance of our findings, we apply our framework to quantum radar and obtain a refined Arthurs-Kelly relation that characterizes the ultimate performance for the simultaneous estimation of range and velocity with any given amount of entanglement. This showcases the transformative potential of our findings for a wide range of applications in quantum metrology, sensing, and beyond. 

	\end{abstract}
 \maketitle

Quantum metrology exploits quantum mechanical phenomena, such as superposition and entanglement, to achieve precision surpassing the limits of classical techniques. While the ultimate bounds for single-parameter estimation are well established through the quantum Cramér-Rao bound (QCRB)~\cite{Hole82book,Hels76book,BrauC94,BrauCM96,Qiushi2023,Kurdzialek22}, the extension to multiparameter scenarios presents significantly greater theoretical and practical challenges~\cite{Szczykulska2016,Rafal2020,ALBARELLI2020126311,vidrighin2014,crowley2014,Yue2014,Zhang2014,Suzuki2016,Liu_2019,ChenHZ2019,Hou20minimal,HouSuper2021,Chen_2017,Luca2017,Kok2020,Liu2017,Roccia_2017,e22111197,singularqfim1,singularqfim2,singularqfim3,singularqfim4,singularqfim5}.
The core difficulty stems from the incompatibility between optimal measurement strategies for different parameters—a uniquely quantum phenomenon with no classical counterpart. This incompatibility arises directly from the non-commutativity of quantum observables, which prevents their simultaneous perfect measurement and gives rise to inherent precision trade-offs in multiparameter estimation. These trade-offs create complex optimization landscapes where improving precision in one parameter inevitably affects others. The development of novel approaches to characterize and quantify these trade-offs is one of the central focuses in quantum metrology~\cite{HongzhenPra,HongzhenPRL,GillM00,HayaM05,Zhu2018universally,Lu2021,Sidhu2021,Nagaoka1,Nagaoka2,Conlon2021,Ragy2016,Federico2021,Carollo_2019,Ragy2016,Candeloro_2021,MatsumotoThesis,Matsumoto_2002,Koichi2013,Kahn2009,Yuxiang2019,FrancescoPRL,Conlon2023,chen2024simultaneous}. 
Prior studies identify the Holevo bound and the Nagaoka–Hayashi bound as tight benchmarks for pure-state models in the general case. However, both typically require numerical evaluation and do not provide an explicit analytical route to constructing the optimal measurement~\cite{Hole82book,Koichi2013,Kahn2009,Yuxiang2019,MatsumotoThesis,Matsumoto_2002,FrancescoPRL,Suzuki2016,Sidhu2021,Nagaoka1,Nagaoka2,Conlon2021}. 
A key gap thus remains in multiparameter quantum estimation: a generally tight analytical bound that captures the trade-off, together with a constructive procedure for designing the measurements saturating the bound.
 
In this Letter, we present a tight analytical trade-off relation that defines the ultimate precision limits for estimating an arbitrary number of parameters encoded in pure quantum states, where measurement incompatibility imposes fundamental constraints. Furthermore, we provide systematic constructions of infinitely many optimal measurements that saturate the trade-off relation. This abundance of optimal measurements is particularly valuable for practical implementations, as it enables the selection of measurement schemes that are most robust against specific noise profiles in a practical scenario. 
We demonstrate the power of our approach by solving a fundamental problem in quantum radar: determining the optimal precision for the simultaneous estimation of range and velocity using entangled light. We derive a refined Arthurs-Kelly uncertainty relation for this scenario, providing a general solution to this longstanding problem in quantum radar. Our approach provides fundamental insights into the trade-offs inherent in multiparameter estimation while simultaneously offering optimal designs of measurement protocols that approach the fundamental limits.

To estimate a set of parameters, $x=(x_1,\cdots, x_n)$, encoded in a quantum state, $\rho_x$, we need to first perform a positive operator-valued measurement (POVM), denoted by $\{M_m\geq 0|\sum_m M_m=I\}$, on the state. The probability of obtaining the measurement result $m$ is given by $p(m|x)=\text{Tr}(M_m\rho_x)$, and from the measurement outcomes, estimators $\hat{x}=(\hat{x}_1,\cdots,\hat{x}_n)$ can be constructed. For any locally unbiased estimators, the covariance matrix is lower bounded as
\begin{equation}\label{eq:QCRB}
    \text{Cov}(\hat{x})\geq \frac{1}{\nu} F_C^{-1}\geq \frac{1}{\nu} F_Q^{-1}.
\end{equation}
The first inequality is the Cramér-Rao bound (CRB)\cite{Cram46}, where $F_C$ is the classical Fisher information matrix (CFIM) with the $jk$-th entry given by $(F_C)_{jk}=\sum_{m}\frac{\partial_{x_j}p(m|x)\partial_{x_k}p(m|x)}{p(m|x)}$ and $\nu$ is the number of repeated measurements. The Cramér-Rao bound can be asymptotically saturated with the maximum-likelihood estimators. Regardless of the choice of the measurement, the CFIM is upper bounded by the quantum Fisher information matrix (QFIM) with $F_C \leq F_Q$, leading to the second inequality in Eq.~(\ref{eq:QCRB}), known as the quantum Cramér-Rao bound (QCRB)\cite{Hels76book}. 
Here $(F_Q)_{jk}=\frac{1}{2}\text{Tr}(\rho_x\{L_j,L_k\})$, where $L_{q}$ is the symmetric logarithmic derivative (SLD) for $x_{q}$, satisfying $\partial_{x_{q}}\rho_x=\frac{1}{2}(\rho_xL_{q}+L_{q}\rho_x)$. In this Letter, we focus on the case that all parameters are locally identifiable where $F_Q$ is invertible.

The QFIM corresponds to the real part of the geometrical tensor $F$, which is an $n\times n$ matrix with $jk$-th entry given by $F_{jk}=\operatorname{Tr}(\rho_x L_jL_k)$. The imaginary part of this tensor, denoted as $F_{\text{Im}}$ with $(F_{\text{Im}})_{jk}=\frac{1}{2i}\text{Tr}(\rho_x[L_j,L_k])$, is proportional to the Berry curvature of the parametrized quantum states, yielding the decomposition $F=F_Q+iF_{\text{Im}}$. Importantly, $F_{\text{Im}}$ provides a quantitative signature of measurement incompatibility in multiparameter estimation. In single-parameter quantum estimation, $F_{\text{Im}}$ naturally vanishes, and an optimal measurement strategy—such as projective measurements on the eigenvectors of the SLD operator—always exists to saturate the QCRB~\cite{BrauC94,BrauCM96}.  In contrast, multiparameter estimation is significantly more complex. A nonzero $F_{\text{Im}}$ indicates the incompatibility of the optimal measurements for different parameters, rendering the QCRB generally unattainable. This leads to a fundamental gap between the classical and quantum Fisher information matrices, resulting in trade-offs in the achievable precision for estimating different parameters~\cite{arthurs1965,Arthurs1988,Ozawa2003,OZAWA2004367,OZAWA2004350,OZAWA200321,Ozawa_2014,Hall2004,Branciard2013,Branciard2014,2014Error,Lu2014,chen2024simultaneous}. 
Quantifying such trade-offs constitutes a central challenge in determining the fundamental precision limits of multiparameter quantum metrology.

To quantify the precision trade-off under general POVMs $\{M_m\}$, we adopt the metric~\cite{MatsumotoThesis,HongzhenPra,HongzhenPRL,GillM00,HayaM05,Zhu2018universally} 
\begin{equation}
    \Gamma=\max_{\{M_m\}}\text{Tr}(F_Q^{-1}F_C),
\end{equation}
which is invariant under reparameterization. When there is no incompatibility, i.e., there exists a POVM to make $F_C=F_Q$, we have $\Gamma=\text{Tr}(I_n)=n$, where $n$ is the number of parameters. In the presence of incompatibility,  $F_C<F_Q$, we have $\Gamma<n$. The gap between $\Gamma$ and $n$ quantifies the trade-off.

As one of the main results of this Letter, we provide a tight analytical bound on $\Gamma$ for the estimation of an arbitrary number of parameters encoded in pure states (see Sec III in \cite{pra} for detailed derivation).
\begin{theorem}\label{theorem}
For the simultaneous estimation of arbitrary $n$ parameters encoded in a pure state $\rho_x=\ket{\Psi_x}\bra{\Psi_x}$, we have
\begin{equation}\label{eq:boundgamma}
\Gamma\leq n-\frac{1}{2}\sum_{q=1}^n (1-\sqrt{1-|\lambda_q|^2}),  
\end{equation}
where $\{\lambda_1,\cdots, \lambda_n\}$ are eigenvalues of $F_Q^{-\frac{1}{2}}F_{\text{Im}}F_Q^{-\frac{1}{2}}$. 
\end{theorem}
The gap between $\Gamma$ and $n$, which is $\frac{1}{2}\sum_{q=1}^n (1-\sqrt{1-|\lambda_q|^2})$, quantifies the precision trade-off imposed by measurement incompatibility. This bound is tight for pure states and also valid for mixed states. 
For mixed states, though it is generally not achievable, the bound can be tighter than existing analytical bounds~\cite{HongzhenPra,HongzhenPRL,GillM00}.

The result includes several known results as special cases. First, the gap vanishes if and only if $F_{\text{Im}}=0$. This is exactly the weak commutativity condition, $\text{Tr}(\rho_x[L_j,L_k])=0$, $\forall j, k$, which is necessary and sufficient for the saturation of QCRB in the case of pure states~\cite{Hole82book,Matsumoto_2002,Luca2017}. Second, for the estimation of $2d-2$ parameters encoded in a $d$-dimensional pure state, all eigenvalues of $F_Q^{-\frac{1}{2}}F_{\text{Im}}F_Q^{-\frac{1}{2}}$ satisfy $|\lambda_q|=1$, $\forall q=1,\cdots, 2d-2$ (see~\cite{pra} for derivation). In this case the trade-off relation reduces to $\text{Tr}(F_Q^{-1}F_C)\leq 2d-2-\frac{1}{2}(2d-2)=d-1$, which recovers the Gill-Massar bound~\cite{GillM00}. For $n<2d-2$, the bound derived here is in general tighter than the Gill-Massar bound~\cite{GillM00}. The detailed derivation of the bound and extensive comparisons with existing bounds can be found in the accompanying paper \cite{pra}.

The trade-off emerges from the fundamental impossibility of simultaneously measuring the non-commuting SLDs, $\{L_j|j=1,\cdots, n\}$, where $L_j$ is associated with parameter $x_j$ and represents the optimal observable for estimating that parameter. 
Quantitatively, this limitation manifests as residual errors when a single POVM is used to approximately measure all the non-commuting $L_j$.
The detailed derivation of this link is presented in Ref.~\cite{pra}. Below, we outline the central concept.
 
A general POVM, $\{M_m\}$, on a state $|\Psi_x\rangle$, can be effectively implemented as a projective measurement, $\{|m\rangle\langle m|\}$, on an augmented state $|\Psi_x\rangle|\xi\rangle$, where $|\xi\rangle$ is an ancillary state. From the projective measurement, we can construct a set of commuting observables, $\{O_j=\sum_m f_j(m)|m\rangle\langle m|| f_j(m)\in \mathbb{R}\}$, to approximate $\{L_j\otimes I\}$, where $I$ is the Identity operator on the ancilla. The mean squared error (MSE) of the approximation  ~\cite{arthurs1965,Arthurs1988,Ozawa2003,OZAWA2004367,OZAWA2004350,OZAWA200321,Ozawa_2014,Hall2004,Branciard2013,Branciard2014,2014Error,Lu2014}
 \begin{eqnarray}\label{eq:error}
 \aligned
\epsilon_j^2&=\bra{\xi}\bra{\Psi_x}(O_j-L_j\otimes I)^2\ket{\Psi_x}\ket{\xi},
 \endaligned
 \end{eqnarray}
is closely related to the discrepancy between $F_C$ and $F_Q$. Specifically, by choosing the optimal $f_j(m)$ under the measurement,
we have $\sum_{j=1}^n \epsilon_j^2=\text{Tr}(F_Q-F_C)$ \cite{pra}.
Under the reparametrization $\tilde{x}=F_Q^{\frac{1}{2}} x$, where the quantum Fisher information matrix changes to $\tilde{F}_Q=I$ and the classical Fisher information matrix changes to $\tilde{F}_C=F_Q^{-\frac{1}{2}}F_CF_Q^{-\frac{1}{2}}$, this yields the relation  
\begin{equation}\label{eq:error_link}
\sum_{j=1}^n \tilde{\epsilon}_j^2=\text{Tr}(\tilde{F}_Q-\tilde{F}_C)=n-\text{Tr}(F_Q^{-1}F_C),
\end{equation}
which directly links the total approximation error to the gap between $n$ and $\text{Tr}(F_Q^{-1}F_C)$. Identifying the optimal set of commuting observables that approximate the SLDs with minimal error corresponds exactly to the minimization of the trade-off.

The total mean squared error can be expressed as a sum of squared Euclidean distances between vectors
\begin{equation}\label{eq:distance}
\sum_{j=1}^n\epsilon_j^2=\sum_{j=1}^n\||o_j\rangle-|l_j\rangle\|^2,    \end{equation}
where $|l_j\rangle=L_j\otimes I|\Psi_x\rangle|\xi\rangle$,
$|o_j\rangle=O_j|\Psi_x\rangle|\xi\rangle$, and $\||v\rangle\|^2=\langle v|v\rangle$. 
Since $[O_j,O_k]=0$, we have $\langle o_j|o_k\rangle=\langle o_k|o_j\rangle$, i.e., $\langle o_j|o_k\rangle\in \mathbb{R}$, $\forall j, k$. 
In particular, in the measurement basis, the observables $O_j=\sum_m f_j(m)|m\rangle\langle m|$ with $f_j(m)\in \mathbb{R}$ are diagonal with real diagonal entries. By appropriately choosing the global phases of the measurement basis $\{|m\rangle\}$, the state \( |\Psi_x\rangle|\xi\rangle \) can also be represented as a real vector in this basis~\cite{note1}, i.e., $|\Psi_x\rangle|\xi\rangle = \sum_m c_m|m\rangle$ with $c_m \in \mathbb{R}$. 
Under this representation, we have $|o_j\rangle=\sum_m f_j(m)|m\rangle\langle m|\Psi_x\rangle|\xi\rangle=\sum_m f_j(m)c_m|m\rangle$ with $f_j(m)c_m\in \mathbb{R}$. 
 
\begin{figure}
  \centering
\includegraphics[width=0.42\textwidth]{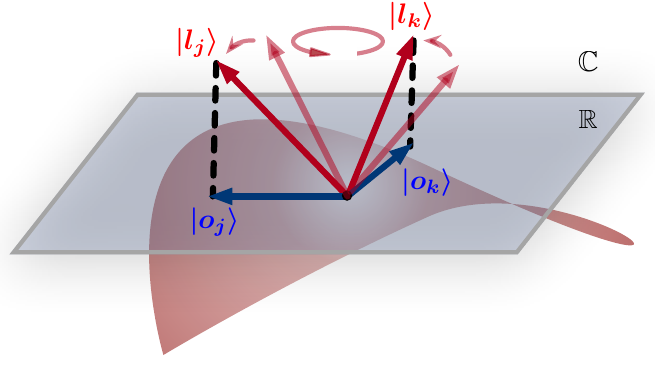}
  \caption{Geometrical picture of the measurement incompatibility. In the measurement basis $\{\ket{m}\}$, $\{\ket{o_j}\}$ are all real vectors while the vectors $\{\ket{l_j}\}$ are in general complex. 
  The incompatibility-induced trade-off is quantified by the Euclidean distances $\sum_j\||o_j\rangle-|l_j\rangle\|^2$. Optimizing the measurement corresponds to applying a unitary rotation to minimize these distances. Note that when $\operatorname{Im}\inp{l_j}{l_k}\neq 0$, no unitary rotation can make $\ket{l_j}$ and $\ket{l_k}$ real vectors at the same time, reflecting the intrinsic incompatibility.
  }
  \label{geo}
\end{figure}

The vectors $|l_j\rangle$, however, are generally complex. We thus have $\||o_j\rangle-|l_j\rangle\|\geq \|\operatorname{Im}|l_j\rangle\|$, where the equality is achieved when $|o_j\rangle=\operatorname{Re}|l_j\rangle$, here $\operatorname{Re}|l_j\rangle$ and $\operatorname{Im}|l_j\rangle$ denote the real and imaginary parts of $|l_j\rangle$, respectively. Finding the optimal measurement thus amounts to identifying a basis $\{\ket{m}\}$ that minimizes $\sum_{j=1}^n \|\operatorname{Im}|l_j\rangle\|^2$. 
Since a change of basis corresponds to applying a unitary rotation to the vectors, the problem reduces to finding the optimal unitary $U$ that minimizes $\sum_{j=1}^n \|\operatorname{Im}(U|l_j\rangle)\|^2$. Note that unitary rotations do not change the inner product of vectors. Thus if $\operatorname{Im}\langle l_j|l_k\rangle\neq 0$, $\operatorname{Im}(U|l_j\rangle)$ and $\operatorname{Im}(U|l_k\rangle)$ can not be simultaneously zero. This yields the geometric picture shown in  Fig.~\ref{geo}:  measurement incompatibility arises from the inability to rotate all vectors $\{\ket{l_j}\}$ simultaneously into the real subspace, and the optimal measurement that minimizes the trade-off rotates them as closely as possible to the real vectors.

We now present a systematic procedure for constructing optimal measurements that saturate the bound in Eq.~(\ref{eq:boundgamma}).
The optimal measurements are derived from $|\Psi_x\rangle|\xi\rangle$ and the optimal $\{|o_j\rangle\}$. 
The explicit analytical forms of optimal $\{|o_j\rangle\}$ that achieve the minimum of Eq.~(\ref{eq:distance}) can be found in Sec.III of the companion paper~\cite{pra}. 
The optimal projective measurement, $\{|m\rangle\}$, corresponds to a basis under which $|\Psi_x\rangle|\xi\rangle$ and the optimal $\{|o_j\rangle\}$ are all real vectors. Such a basis can be identified through the following \emph{two-step procedure}:

(i) First, apply the Gram-Schmidt orthonormalization to the set $\{|\Psi_x\rangle|\xi\rangle, |o_1\rangle, |o_2\rangle, \cdots, |o_{n}\rangle\}$ to obtain an orthonormal set $\{|a_0\rangle, |a_1\rangle, |a_2\rangle, \cdots, |a_{n}\rangle\}$.
These vectors can be expanded into a complete basis by adding additional orthonormal vectors $\{|a_j\rangle| j=n+1,\cdots, d-1\}$, here $d$ is the dimension of the Hilbert space for the system+ancilla. 

(ii) Second, choose a real orthonormal basis, $\{|b_0\rangle,|b_1\rangle,\cdots, |b_{d-1}\rangle\}$, as the representation of $\{|a_0\rangle,|a_1\rangle,\cdots, |a_{d-1}\rangle\}$ in the measurement basis. The change from the original basis to the measurement basis is represented by the unitary that transforms $\{|a_j\rangle\}$ to $\{|b_j\rangle\}$, which is given by $U=BA^{-1}$, where $A=(|a_0\rangle,|a_1\rangle,\cdots, |a_{d-1}\rangle)$ is a matrix whose columns are $\{|a_j\rangle\}$, and $B=(|b_0\rangle,|b_1\rangle,\cdots, |b_{d-1}\rangle)$ is the real orthogonal matrix whose columns are $\{|b_j\rangle\}$. The optimal measurement basis is then given by the rows of $U$. 

The optimal measurement is not unique: different choices of the matrix $B$ lead to different optimal measurements. The only constraint on $B$ is that when the $m$-th entry of the first column of $B$ (i.e., the $m$-th entry of $|b_0\rangle$) is zero, the $m$-th entries of subsequent columns, $|b_j\rangle$ for $j = 1, \cdots, n$, should also be zero~\cite{note2}. In particular, any $B$ with a first column devoid of zero entries is permissible.

Moreover, because the first column of $B$ corresponds to the vector $(\langle 1|\Psi_x\rangle|\xi\rangle,\cdots, \langle m|\Psi_x\rangle|\xi\rangle,\cdots)^T$, by choosing $B$ properly we can control the measurement probabilities $\{ p_m = |\langle m | \Psi_x \rangle | \xi \rangle |^2 \}$. This flexibility enables the engineering of probability distributions that are more robust to practical noise, a feature of significant practical relevance~\cite{kurdzialek2023measurement}. In the companion paper~\cite{pra}, we demonstrate this feature on a cloud superconducting quantum computing platform~\cite{quafu3}.

\begin{figure}[t]
  \centering
  \includegraphics[width=0.5\textwidth]{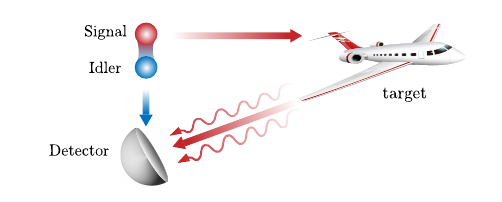}
  \caption{Radar sensing of target's range and velocity using entangled biphoton state.}
  \label{radar}
\end{figure}

We next apply our results to quantum radar, focusing on the simultaneous estimation of the range and velocity of a moving object. 
While precision limits of quantum radar have been explored in depth~\cite{zhuang2017,zhuang2021,zhuang2022,huang2021,quantumradar2020,Maccone2020,reichert2022quantum,reichert2024heisenberg,li2023entanglement}, they are primarily established under extreme assumptions—either separable photons or perfectly entangled biphotons~\cite{zhuang2017,huang2021}.
Despite considerable research efforts, two critical questions remain unresolved: (1) the ultimate precision limit achievable with a practical biphoton source that possesses only a finite amount of entanglement and (2) the corresponding optimal measurement strategies. This knowledge gap persists primarily due to the theoretical challenges posed by measurement incompatibility in such scenarios. Our method provides the necessary tools to address these questions and establish fundamental limits for quantum radar with practical biphoton sources.

Following the treatment of Ref.~\cite{huang2021}, we consider the pulsed quantum radar where entangled biphotons are employed to estimate the range and velocity of a target simultaneously. The signal photon is transmitted toward the target and coherently reflected, while the idler photon is retained and later measured jointly with the returned signal. To establish fundamental benchmarks, we focus on the idealized scenario where the target acts as a perfect mirror and background noise is negligible.

The range and velocity are encoded in the time of flight and Doppler shift of the signal photon, respectively, as illustrated in Fig.\ref{radar}. For Gaussian pulses, the state of the reflected signal and the idler is given by~\cite{huang2021,pra} $\ket{\Psi} = \int dt \int dt_i \Psi(t,t_i) \ket{t} \ket{t_i}$,
where
\begin{equation}
 \begin{aligned}
     \Psi(t,t_i) = &\mathcal{N} \exp\{-i \bar{\omega}(t-\bar{t})-i\bar{\omega}_{i0}(t_i-\bar{t}_{i0})-(t-\bar{t})^2 \sigma^2 \\&- (t_i-\bar{t}_{i0})^2 \sigma_{i0}^2 + 2 \kappa (t-\bar{t})(t_i-\bar{t}_{i0})\sigma\sigma_{i0}\}.
 \end{aligned}
\end{equation}
Here $\sigma = \frac{c-v}{c+v} \sigma_0$, $\bar{t} = \bar{t}_0 + \frac{2x}{c-v}$, $\bar{\omega} = \frac{c-v}{c+v} \bar{\omega}_0$, and the normalization factor $\mathcal{N} = \sqrt{\frac{2\sigma \sigma_{i0}}{\pi}}(1-\kappa^2)^{1/4}$, $\bar{t}_0$($\bar{t}_{i0}$), $\bar{\omega}_0$($\bar{\omega}_{i0}$), $\sigma_0$($\sigma_{i0}$) denote the central time, central frequency, and bandwidth of the sent (idler) photon. The coefficient $\kappa \in [0,1)$ is the correlation between the signal and the idler photon, which quantifies the entanglement between two photons. 
$\kappa=0$ corresponds to two separable photons, while the limit $\kappa \to 1$ corresponds to perfect entanglement, which is not realizable in practice.
The range $x$ and velocity $v$ can be obtained from the central time $\bar{t}$ and the frequency $\bar{\omega}$ of the returned signal photon. The quantum Fisher information matrix for $\bar{t}$ and $\bar{\omega}$ is $F_Q = \left(\begin{matrix}
    4 \sigma^2 & 0 \\ 0 & \frac{1}{\sigma^2(1-\kappa^2)}
  \end{matrix}\right)$.
A trade-off relation for the simultaneous estimation of $\bar{t}$ and $\bar{\omega}$ has previously been derived directly from QCRB ~\cite{huang2021}:
\begin{equation}\label{eq:boundqcrb}
\sigma_{\bar{t}}\sigma_{\bar{\omega}} \geq \frac{\sqrt{1-\kappa^2}}{2},
\end{equation}
here $\sigma_{\bar{t}}$ and $\sigma_{\bar{\omega}}$ denote the standard deviations for the estimation of $\bar{t}$ and $\bar{\omega}$ respectively. 

However, in this case the weak commutativity condition is violated, since 
\begin{equation}
F_{\text{Im}}=\left(\begin{matrix}
     0 & -2 \\ 2 & 0
   \end{matrix}\right)\neq \left(\begin{matrix}
     0 & 0 \\ 0 & 0
   \end{matrix}\right),   
\end{equation}
Hence the QCRB is unattainable, and the bound  in \eqref{eq:boundqcrb} derived directly from it is overly optimistic. 
\begin{figure}[t]
  \centering
  \includegraphics[width=0.48\textwidth]{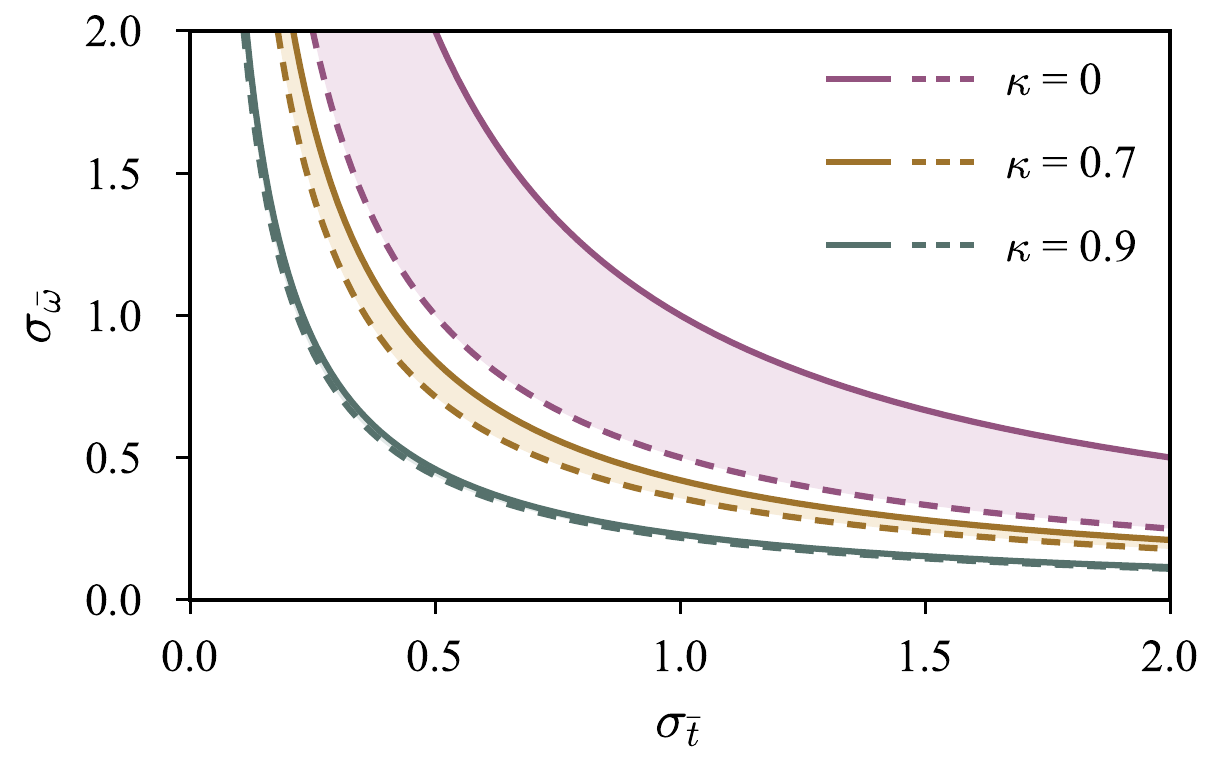}
  \caption{Uncertainty relations in quantum radar using biphoton states with different amounts of entanglement ($\kappa$). Here the dashed lines denote the uncertainty relations obtained from QCRB and the solid lines denote the refined Arthurs-Kelly uncertainty relations, which is achievable for the simultaneous estimation of range and velocity. 
  The shaded areas illustrate the gap between the QCRB and the tight bound.} 
  \label{radar_uncertainty}
\end{figure}

For separable photons ($\kappa=0$), Arthurs and Kelly obtained a tighter, achievable bound $\sigma_{\bar{t}}\sigma_{\bar{\omega}}\geq 1$~\cite{arthurs1965}. For perfectly entangled photons ($\kappa=1$), Zhuang et.al~\cite{zhuang2017} constructed a heuristic measurement that achieves $\sigma_{\bar{t}}\sigma_{\bar{\omega}}=0$. Yet for general $\kappa$ with $0<\kappa<1$, tight trade-off relations remain unknown. 

Our trade-off relation closes this gap. For any $\kappa$, we have $F_Q^{-\frac{1}{2}}F_{\text{Im}}F_Q^{-\frac{1}{2}}=\left(\begin{matrix}
     0 & -\sqrt{1-\kappa^2} \\ \sqrt{1-\kappa^2} & 0
   \end{matrix}\right)$, whose eigenvalues are $\pm i\sqrt{1-\kappa^2}$. Applying the general bound in Eq.~(\ref{eq:boundgamma}) yields
\begin{equation}
    \text{Tr}(F_Q^{-1}F_C) \leq 2 - (1- \kappa) = 1+\kappa ,
\end{equation} 
which can be restated as \[\frac{(F_C)_{11}}{4\sigma^2}+\sigma^2(1-\kappa^2)(F_C)_{22}\leq 1+\kappa.\] 
Using the inequality $\sqrt{1-\kappa^2}\sqrt{(F_C)_{11}(F_C)_{22}}\leq \frac{(F_C)_{11}}{4\sigma^2}+\sigma^2(1-\kappa^2)(F_C)_{22} $ and the fact that $\sigma_{\bar{t}}\sigma_{\bar{\omega}}\geq \frac{1}{\sqrt{(F_C)_{11}(F_C)_{22}}}$, we obtain the tight bound
\begin{equation}
\sigma_{\bar{t}}\sigma_{\bar{\omega}} \geq \frac{\sqrt{1-\kappa^2}}{1+\kappa}=\frac{\sqrt{1-\kappa}}{\sqrt{1+\kappa}}.
\end{equation}
This relation unifies the previous limiting cases: it recovers the Arthurs–Kelly relation at $\kappa=0$ and reaches the ideal entangled limit ($\sigma_{\bar{t}}\sigma_{\bar{\omega}}=0$) at $\kappa=1$. The differences between this relation and Eq.~(\ref{eq:boundqcrb}) derived directly from the QCRB can be found in Fig. \ref{radar_uncertainty}. We can also rigorously confirm now that the heuristic continuous‑spectrum measurement proposed by Zhuang et al.~\cite{zhuang2017,huang2021} is optimal, as it saturates the refined relation. While its optimality was previously known only in the perfect-entanglement limit ($\kappa \to 1$), our formalism extends it to all $\kappa$. More importantly, our general construction procedure reveals that all essential information is encoded in a finite‑dimensional subspace. Consequently, optimal measurements can be implemented as projective measurements with a finite number of outcomes, as we explicitly construct in the companion paper~\cite{pra}. This provides flexibility for tailoring implementations to specific physical platforms where continuous-spectrum detection may be challenging, and discrete, projective measurements are preferred.

\textit{Summary.}---The incompatibility inherent in estimating multiple parameters has posed a significant challenge to determining the ultimate precision limits in multi‑parameter quantum metrology. We have addressed this challenge by deriving a tight analytical trade‑off relation for the estimation of multiple parameters in pure states and by providing systematic constructions of infinitely many optimal measurements that saturate this bound. Our results not only establish a precise, attainable bound but also offer a clearer geometrical picture of incompatibility‑induced trade‑offs, thereby deepening the conceptual understanding of multiparameter quantum estimation. Our method shows significant potential for evaluating the ultimate performance in various applications, such as quantum radar, quantum gyroscopes, and quantum imaging, where simultaneous estimation is essential. By addressing the issue of measurement incompatibility, our work offers a valuable approach for developing more precise and efficient quantum measurement strategies.

\begin{acknowledgments}
This work is supported by the Quantum Science and Technology-National Science and Technology Major Project (2023ZD0300600), the Research Grants Council of Hong Kong (14309223, 14309624, 14309022), Guangdong Provincial Quantum Science Strategic
Initiative (GDZX2303007,GDZX2505003), 1+1+1 CUHK-CUHK(SZ)-GDST Joint Collaboration Fund (Grant No. GRDP2025-022). H. C. acknowledges the support from the National Natural Science Foundation of China (Grants No. 12505024, No. 92476201), Department of Science and Technology of Guangdong Province (Grant No. 2024QN11X234), Guangdong Basic and Applied Basic Research Foundation (Grant No. 2025A1515011441), Shenzhen Science and Technology Program (Grants No. ZDYJ20251211120900001, No. JCYJ20240813141350066).
\end{acknowledgments}


\begin{thebibliography}{99}
\bibitem{note1}
Specifically, expressing \( |\Psi_x\rangle|\xi\rangle = \sum_m c_m|m\rangle \) with \( c_m = \langle m|\Psi_x\rangle|\xi\rangle \), we note that if \( c_m = r_m e^{i\phi_m} \) is complex, we can redefine the basis states as \( |\tilde{m}\rangle = e^{i\phi_m}|m\rangle \), where $\{\ket{\tilde{m}}\bra{\tilde{m}}\}$ corresponds to the same projective measurement as $\{|m\rangle\langle m|\}$ and the state becomes \( |\Psi_x\rangle|\xi\rangle = \sum_m r_m|\tilde{m}\rangle \), where \( r_m \in \mathbb{R} \). Thus, without loss of generality, we may assume \( c_m \in \mathbb{R} \).
\bibitem{note2}
Essentially $|b_0\rangle=(\langle 1|\Psi_x\rangle|\xi\rangle, \cdots, \langle m|\Psi_x\rangle|\xi\rangle, \cdots)^T$ is the representation of $|\Psi_x\rangle|\xi\rangle$ in the measurement basis. If $\langle m|\Psi_x\rangle|\xi\rangle=0$, then in the measurement basis the $m$-th entries of $|o_j\rangle=\sum_m f_j(m)|m\rangle\langle m|\Psi_x\rangle|\xi\rangle$, $j=1,\cdots, n$, should also be zero.
\end{thebibliography}
\end{document}